\documentclass[draft]{agujournal2019}
\usepackage{url} 
\usepackage{lineno}
\usepackage[inline]{trackchanges} 
\usepackage{soul}
\usepackage{graphicx}
\usepackage{subfigure}
\draftfalse

\journalname{Geophysical Research Letters}

\begin{document}

%
%


\title{X-Raying Neutral Density Disturbances in the Mesosphere and Lower Thermosphere induced by the 2022 Hunga-Tonga Volcano Eruption-Explosion}

%
%




\authors{Satoru Katsuda\affil{1}, Hiroyuki Shinagawa\affil{2}, Hitoshi Fujiwara\affil{3}, Hidekatsu Jin\affil{4}, Yasunobu Miyoshi\affil{5}, Yoshizumi Miyoshi\affil{6}, Yuko Motizuki\affil{1,7}, Motoki Nakajima\affil{8}, Kazuhiro Nakazawa\affil{9}, Kumiko K. Nobukawa\affil{10}, Yuichi Otsuka\affil{6}, Atsushi Shinbori\affil{6}, Takuya Sori\affil{6}, Chihiro Tao\affil{4}, Makoto S. Tashiro\affil{1,11}, Yuuki Wada\affil{12}, Takaya Yamawaki\affil{1}}


\affiliation{1}{Graduate School of Science and Engineering, Saitama University, 255 Shimo-Ohkubo, Sakura, Saitama 338-8570, Japan}
\affiliation{2}{International Research Center for Space and Planetary Environmental Science, Kyushu University, Fukuoka 819‑0395, Japan}
\affiliation{3}{Faculty of Science and Technology, Seikei University, Tokyo, Japan}
\affiliation{4}{National Institute of Information and Communications Technology, Koganei, Tokyo 184-8795, Japan}
\affiliation{5}{Department of Earth and Planetary Sciences, Faculty of Science, Kyushu University, Fukuoka 819‑0395, Japan}
\affiliation{6}{Institute for Space‑Earth Environmental Research, Nagoya University, Nagoya, Aichi 464‑8601, Japan}
\affiliation{7}{RIKEN Nishina Center, Hirosawa 2-1, Wako 351-0198, Japan}
\affiliation{8}{School of Dentistry at Matsudo, Nihon University, 2-870-1 Sakaecho-nishi, Matsudo, Chiba 101-8308, Japan}
\affiliation{9}{Kobayashi-Maskawa Institute for the Origin of Particles and the Universe, Nagoya University, Furo-cho, Chikusa-ku, Nagoya, Aichi 464-8601, Japan}
\affiliation{10}{Faculty of Science and Engineering, Kindai University, 3-4-1 Kowakae, Higashi-Osaka, 577-8502, Japan}
\affiliation{11}{Institute of Space and Astronautical Science (ISAS), Japan Aerospace Exploration Agency (JAXA), 3-1-1 Yoshinodai, Chuo-ku, Sagamihara, Kanagawa, 252-5210, Japan}
\affiliation{12}{Graduate School of Engineering, Osaka University, 2-1 Yamadaoka, 565-0871, Suita, Osaka, Japan}



\correspondingauthor{Satoru Katsuda}{katsuda@mail.saitama-u.ac.jp}




\begin{keypoints}
\item Neutral density profiles in the MLT were measured before and after Tonga's huge volcanic eruption on 15 January 2022.  
\item Shortly after the eruption, a strong and long-lasting neutral density depletion was found near the epicenter.  
\item Density profiles after the eruption showed wavy structures with a typical wavelength of either $\sim$20\,km (vertical) or $\sim$1,000\,km (horizontal).  
\end{keypoints}

%
%

%
%


\begin{abstract}
We present X-ray observations of the upper atmospheric density disturbance caused by the explosive eruption of the Hunga Tonga-Hunga Ha'apai (HTHH) volcano on 15 January 2022.  From 14 January to 16 January, the Chinese X-ray astronomy satellite, Insight-HXMT, was observing the supernova remnant Cassiopeia~A.  The X-ray data obtained during Earth's atmospheric occultations allowed us to measure neutral densities in the altitude range of $\sim$90--150\,km.  The density profiles above 110\,km altitude obtained before the major eruption are in reasonable agreement with expectations by both GAIA and NRLMSIS~2.0 models.  In contrast, after the HTHH eruption, a severe density depletion was found up to at least 1,000\,km away from the epicenter, and a relatively weak depletion extending up to $\sim7,000$\,km for over 8\,hr after the eruption.  In addition, density profiles showed wavy structures with a typical length scale of either $\sim$20\,km (vertical) or $\sim$1,000\,km (horizontal).  This may be caused by Lamb waves or gravity waves triggered by the volcanic eruption.
\end{abstract}

\section*{Plain Language Summary}
Volcanic eruptions trigger acoustic and gravity waves that propagate vertically upward and cause significant perturbations in the ionosphere.  Lots of observations revealed ionospheric disturbances.  However, there have been limited observations of the mesosphere and the lower thermosphere (MLT) --- an important atmospheric layer that connect the ionospheric disturbances and the volcanic eruption.  On 15 January 2022, a huge explosive eruption occurred at Hunga Tonga-Hunga Ha'apai (HTHH), providing us with excellent opportunities to study the atmosphere-ionosphere disturbances driven by volcanic eruptions.  Here, we first reveal MLT neutral density disturbances caused by the HTHH eruption, based on atmospheric occultations of the celestial X-ray bright source, Cassiopeia~A, observed with the Chinese X-ray astronomy satellite Insight-HXMT.  Shortly after the HTHH eruption, we found severe density depletions in the vicinity of the epicenter, by factors of $\sim$2--10 less than the control experiments by two atmospheric models.  The strongest density depletion was found at 500--1,000\,km away from the epicenter, and a relatively weak depletion extends up to $\sim$7,000\,km for at least 8\,hr after the eruption.  This density behavior is qualitatively consistent with that seen at a much higher altitude of $\sim$500\,km.  In addition, neutral density profiles obtained in the MLT suggest a typical length scale of either $\sim$20\,km (vertical) or $\sim$1,000\,km (horizontal).  This may be caused by Lamb waves or gravity waves triggered by the volcanic eruption.  This study demonstrates the power of the X-ray remote sensing technique to investigate little-known behaviors of the MLT.

%
%

\section{Introduction}

On 15 January 2022 at about 4:15 UT, a huge explosive eruption occurred at Hunga Tonga-Hunga Ha'apai, hereafter HTHH located at 20.5$^\circ$S, 184.6$^\circ$E.  The magnitude of the eruption is comparable with and seems the largest after the 1883 eruption in Krakatau.  Thanks to its remarkable scale as well as the state-of-the-art observation instruments, the HTHH eruption provided us with excellent opportunities to study impacts of the volcanic eruption on the atmosphere in unprecedented detail.  In fact, a lot of observations have been revealing ionospheric disturbances, including traveling ionospheric disturbances, occurrence of plasma bubbles, and a significant depletion in total electron content (TEC) above the eruption region \cite<e.g.,>{2023EPS...75..175S,2024EPS...76...15S}.  The atmospheric response to the HTHH eruption can be divided into two types: direct explosive effects near the volcano and the effects caused by a Lamb wave due to the volcanic eruption in the far field.  These ionospheric disturbances are thought to be caused by lower atmospheric forcing.  However, no direct connection between the ionospheric disturbances and the volcanic eruption has been established, mainly because observations are scarce in the mesosphere and the lower thermosphere (MLT).  Bridging this observational gap is essential to fully understand the volcanic impacts on the upper atmosphere.  

Although signatures of the volcanic eruption in the MLT have been a challenge to identify, there are some interesting observational results.  \citeA{2022Natur.609..741W} found Lamb waves concentrically propagating from the HTHH volcano, traveling with a typical speed of 318\,m\,s$^{-1}$.  They reported that the Lamb wave was seen 4.3\,hr after the eruption, vertically extending up to $\sim$87\,km altitude as phase fronts in hydroxyl airglow over Hawaii with a distance of 4,960\,km away from the epicenter.  On the other hand, using the Chilean Observation Network De Meteor Radars (CONDOR) and Nordic Meteor Radar Cluster located at $\sim$10,500\,km and $\sim$15,000\,km away from the HTHH volcano, respectively, \citeA{2023AnGeo..41..197S} concluded that the primary Lamb waves caused by the volcanic eruption did not reach the upper atmosphere in 80--100\,km.  Instead, they identified volcano-triggered gravity waves in the MLT winds with a phase speed of $\sim$200\,m\,s$^{-1}$ and a horizontal wavelength of 1,600--2,000\,km.  Another interesting horizontal wind was found in the altitude range of 80--100\,km on the west side of South America spanning more than 3,000\,km, based on three multistatic specular meteor radars \cite{2023GeoRL..5003809P}.  The wave is characterized as traveling at a phase speed of $\sim$200\,m\,s$^{-1}$, with a period of $\sim$2\,hr and a horizontal wavelength of $\sim$1,440 km in the longitudinal direction, and was interpreted as a high-order Lamb wave.  We note that some of the observed Traveling Ionospheric Disturbances have similar velocities to those of the Lamb waves \cite{2022JGRA..12730735M,2023GeoRL..5003809P,2023EP&S...75...92T}.


In addition to the wind observations, \citeA{2022GeoRL..4998339L} identified strong and slow gravity waves in the altitude range of 50--100\,km by the vertical temperature profiles obtained with TIMED/SABER.  Their vertical and horizontal wavelengths vary in the ranges of 13.9--25.5\,km and 163--857\,km with group velocities of 0.8--4.4\,m\,s$^{-1}$ and 44--81\,m\,s$^{-1}$, respectively.  Their wavelengths and velocities are comparable with usual gravity waves in the MLT, but their amplitudes are enhanced than usual.  The Global-scale Observations of the Limb and Disk (GOLD) mission observed global-scale neutral temperature perturbations (deviations $\sim100$\,K) at $\sim$150\,km altitude from 12 to 16 UT \cite{2023GeoRL..5003158A}.  This temperature perturbation is initially roughly coincident and morphologically similar to the tropospheric pressure-wave structure \cite{2022GeoRL..4998240A}, suggesting that the temperature perturbation is caused by eruption-induced atmospheric waves.  The column O/N$_2$ composition change measured with GOLD seem to be dominated by effects of a geomagnetic storm that occurred at 23 UT on 14 January; strong O/N$_2$ enhancement at low latitudes and depletion at high latitudes \cite{2023GeoRL..5003158A}.  However, less significant O/N$_2$ enhancements are seen toward the HTHH volcano.  This seems consistent with a long-lasting O/N$_2$ depletion near the eruption site observed with TIMED/GUVI \cite{2023JGRA..12830984H}.

Here, we present the first direct measurement of neutral density variations in the MLT between 90 and 150\,km altitudes before and after the HTHH eruption.  We realize an X-ray tomography of the atmosphere using X-ray astronomy satellites.  For the HTHH eruption event, we analyze atmospheric X-ray occultations of the bright supernova remnant, Cassiopeia~A, observed with the Chinese X-ray astronomy satellite Insight-HXMT.  The X-ray occultation method allows us to measure combined atomic and molecular number densities integrated along the line of sight, i.e., the column densities, as has been demonstrated in the literature \cite{2007JGRA..112.6323D,2021JGRA..12628886K,2022AMT....15.3141Y,2023JGRA..12830797K,2023ApJS..264....5X}.  This paper is organized as follows.  In Section 2, we describe observations analyzed in this paper.  In Section 3, we describe data analysis and results.  In Section 4, we interpret the analysis results. Finally, we give conclusions of this paper in Section 5.

\section{Observations and Data Reduction}

As with many X-ray astronomy satellites, the Chinese X-ray astronomy satellite Insight-HXMT \cite{2020SCPMA..6349502Z} is in a 550\,km low-earth orbit.  When Insight-HXMT is used in a pointing mode, observations are conducted with a fixed attitude, with a direction of the X-ray telescope being fixed to a celestial source during each observation.  Therefore, we expect a pair of setting and rising of the X-ray source every orbit.  Atmospheric X-ray occultations occur just before and after the setting and rising.  We can measure density profiles from gradual decrease/increase of the X-ray intensity while the tangent altitude is between $\sim$50\,km and $\sim$200\,km; here, the tangent altitude is the height from the Earth surface at the tangent point --- the location at which the line of sight is closest to the Earth.  Since higher-energy X-rays can penetrate the atmosphere more deeply, the altitude range that can be investigated with this method is sensitive to both the spectral hardness (or mean photon energy) of the X-ray source and the energy coverage of the instrument; the higher the X-ray energy, the deeper the atmospheric layer investigated.  We also note that speeds of tangent-altitude movements during occultations range from $\sim$2\,km\,s$^{-1}$ for a normal (vertical) occultation to $\sim$0.1\,km\,s$^{-1}$ for a grazing (quasi-parallel) occultation.  

Insight-HXMT was observing the bright supernova remnant, Cassiopeia~A (hereafter, Cas~A), from 14 January at 1.5 UT to 16 January at 1.5 UT in 2022.  Cas~A is one of the youngest supernova remnants in our Galaxy with the estimated date of the supernova explosion being AD~1681$\pm$19 \cite{2006ApJ...645..283F}.  Because Cas~A is very bright and stable, it can be used as a background (BG) source for the atmospheric occultation.  During this Cas~A observation, we obtained a total of 15 clean atmospheric occultations for our analyses as listed in Table~\ref{tab:obs_list}.  Figure~\ref{fig:lospos_densityR} upper panel shows trajectories of tangent points during atmospheric occultations (tangent altitudes within 75--150\,km) overlaid on the Earth's surface map.  

Occultation IDs 11--15 in Table~\ref{tab:obs_list} are grazing, i.e., Cas~A was not completely occulted by the Earth's atmosphere.  Because of the grazing condition, the low ends of altitude ranges that can be investigated are limited to 94\,km, 94\,km, 100\,km, 110\,km, and 115\,km for occultation IDs~11, 12, 13, 14, and 15, respectively.  The lowest tangent altitudes reached by other occultations are below $\sim$90\,km, and thus their low-end altitudes are naturally limited to $\sim$90\,km due to the complete atmospheric X-ray absorption of Cas~A with Insight-HXMT/LE.  

Given that Cas~A is a relatively soft X-ray source compared with the Crab Nebula, which has been used as a background source for this method \cite{2007JGRA..112.6323D,2021JGRA..12628886K,2022AMT....15.3141Y,2023JGRA..12830797K,2023ApJS..264....5X}, we here focus on data acquired with the Low Energy X-ray telescope (LE: \citeA{2020SCPMA..6349505C}).  The energy range covered with LE is 1--10\,keV, which is sensitive to atmospheric absorption in the altitude range of $\sim$90--150\,km.  We used the Insight-HXMT Data Analysis Software package4 (HXMTDAS) of version 2.05 and the CALibration DataBase (CALDB) of version 2.06 to extract high-level scientific products.  We filtered the data using standard criteria, except for occultation IDs~10, 11, and 13, for which we set the cut-off-rigidity (COR) condition to ``COR $>$ 5, 4, 6", respectively, so that we can obtain occultation data from them.


\section{Analysis and Results}

\begin{figure}[t]
\begin{tabular}{cc}
\centering
\includegraphics[scale=1.0]{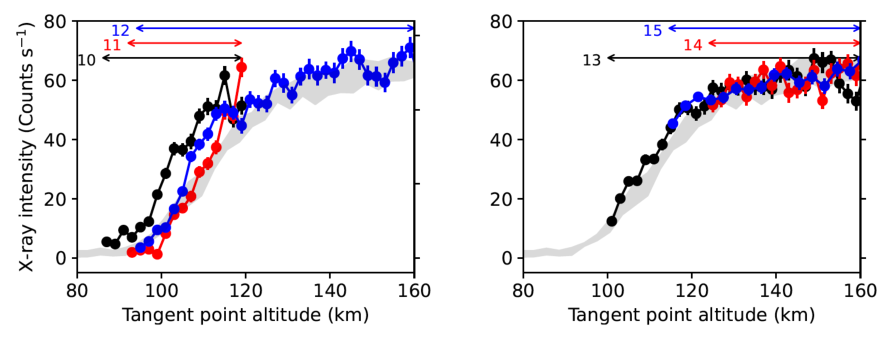}
\end{tabular}
\caption{Left: Occultation light curves in the 1.3--7.3\,keV band for occultation IDs~10 (black), 11 (red), 12 (blue), and the standard deviation of IDs~1--9 (gray).  The altitude range covered by each data is indicated as arrows associated with occultation IDs.  Right: Same as left but for IDs~13 (black), 14 (red), 15 (blue).}
\label{fig:LC}
\end{figure}

We first checked light curves for the 15 occultations obtained with Insight-HXMT/LE.  Figure~\ref{fig:LC} shows BG-subtracted X-ray intensities as a function of tangent altitude.  Since Insight-HXMT/LE is a non-imaging instrument, we cannot subtract BGs from the off-source region in the same field of view.  To subtract the BG emission in our light curve analysis, we generated two light curves for each occultation in two energy bands: 1.3--7.3\,keV and 10.0--12.5\,keV for source and BG, respectively.  Then, we generated a BG-subtracted light curve, by subtracting the BG light curve from the source.  Here, we scaled the BG light curve by multiplying by a factor of 0.45, which was evaluated by an X-ray spectrum in 0--80\,km altitudes integrated over occultation IDs~1--9.  This scaling factor could vary by a factor of 2, depending on the COR condition.  However, given that the BG rate is generally as small as $\sim$1.5\,cts\,s$^{-1}$ in the 1.3--7.3\,keV, the scaling factor does not affect the light curve above 100\,km altitude.  The gray area in Fig.~\ref{fig:LC} represents the standard deviation around the average of the 9 occultation light curves taken on 14 January (occultation IDs~1--9), whereas other data with filled circles are responsible for occultation IDs~10, 11, and 12 (left panel) and 13, 14, and 15 (right panel).  The atmospheric absorption is evident below $\sim$130\,km altitude.  By comparing the pre- and post-eruption X-ray intensities at the same altitude, we can see that post-eruption intensities are stronger than pre-eruption.  This implies less absorption by the atmosphere after the eruption, which is likely caused by a density depletion after the eruption.  Given the exponential vertical density profile, most of the X-ray emission is absorbed by the atmosphere near the tangent point.  Therefore, it is reasonable to attribute the less X-ray absorption in the altitude range of 90--120\,km to the atmospheric density depletion there.  Another notable feature in Fig.~\ref{fig:LC} is the wavy structure of the X-ray intensity.  For example, the X-ray intensity of occultation ID~11 is significantly weaker and stronger at altitudes of $\sim$100\,km and 120\,km, respectively, than the standard deviation of pre-eruption data.  

\begin{figure}[t]
\begin{tabular}{cc}
\centering
\includegraphics[scale=1.0]{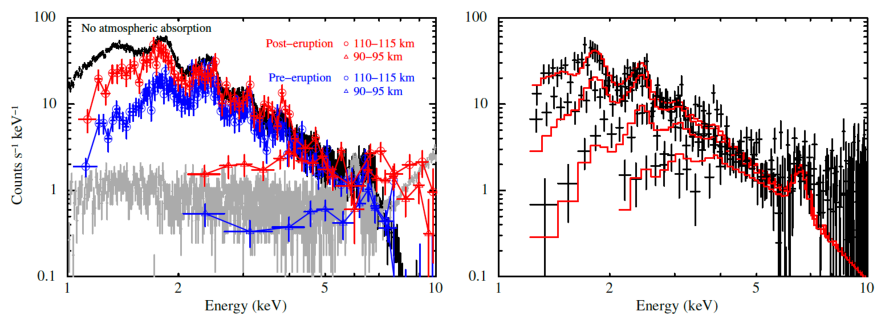}
\end{tabular}
\caption{Left: X-ray spectral comparison before (blue) and after (red) the HTHH eruption on 15 January 2022.  The data with open circles and triangles are responsible for altitude ranges of 110--115\,km and 90--95\,km, respectively.  For comparison, a typical Cas~A spectrum obtained during negligible atmospheric absorption is shown in black, for which a BG spectrum scaled to match the count rate in 10.5--12.5\,keV is shown in gray.  Right: X-ray spectral variation during the atmospheric occultation.  The four datasets are responsible for tangential altitude ranges of 110--115\,km 100-105\,km, 95--100\,km, and 90--95\,km, respectively.  The best-fit models are shown in solid red lines.}
\label{fig:spec}
\end{figure}

Figure~\ref{fig:spec} left and right demonstrate spectral variations during occultations obtained with Insight-HXMT/LE.  The X-ray spectrum is dominated by line emission, which arises from thermal X-ray emission from hot plasmas with multiple electron temperatures of a few 10$^7$\,K.  In the left panel, the black spectrum is taken during negligible atmospheric absorption.  The two blue spectra with open circles and triangles are pre-eruption data integrated over IDs~1--9 in altitude ranges of 110--115\,km and 90--95\,km, respectively.  As expected, low-energy X-rays decrease with decreasing tangential altitude.  The two red data are responsible for the post-eruption data, i.e., occultation ID~10 taking place 3.4\,hr after the major eruption.  They are more absorbed than the pre-eruption data when compared at the same altitude ranges.  This spectral difference is direct evidence that the spectral change before and after the HTHH eruption is caused by a decreased atmospheric absorption after the eruptive explosion.  

As BG for our spectral analysis, we a used pre-eruption spectrum in 0--80\,km altitudes integrated over occultation IDs~1--9.  This BG spectrum was scaled to match the count rate in 10.5--12.5\,keV of source spectra.  As explained above, the BG spectral shape depends on the COR.  However, we checked that the BG spectrum with different COR values do not affect the fitting results.

Next, we measured density profiles for each data group defined in Table~\ref{tab:obs_list}, by analyzing X-ray spectra sliced by altitude.  The effective exposure time for each occultation depends on the duration of the tangent altitude within $\sim$90--150\,km.  Looking at Fig.~\ref{fig:lospos_densityR} upper panel, occultations on 14 January have relatively shorter exposure times than those on 15 January.  Therefore, the pre-eruption data are combined by two or three to improve the number of X-ray photons and reduce the statistical uncertainties on our density measurements.  In the following spectral analysis, we used the XSPEC package \cite{1996ASPC..101...17A}.  We first determined the unattenuated spectral model for each data group.  To this end, we employed a phenomenological model consisting of absorbed (TBabs in XSPEC), bremsstrahlung and power-law for continuum emission and 10 Gaussians for line emission, where the absorption component represents the interstellar absorption between Cas~A and the Earth, i.e., TBabs$\times$(bremss + power-law + Gaussian + Gaussian + ... + Gaussian).  Then, we fitted lower-altitude X-ray spectra with a spectral model taking account of the atmospheric absorption, for which we considered both the photo-electric absorption (vphabs in XSPEC) and the Compton scattering (cabs in XSPEC) due to the Earth's atmosphere , i.e., vphabs$\times$cabs$\times$TBabs$\times$(bremss + power-law + Gaussian + Gaussian + ... + Gaussian).  In this fitting process, we fixed all the spectral parameters determined above, except for the N column density in the atmospheric absorption component.  To mimic the Earth's atmosphere, we fixed the H column density in the vphabs component to a negligible value of 10$^{12}$\,atoms\,cm$^{-2}$.  The current X-ray data do not allow us to measure the relative abundances among N, O, and Ar in the atmosphere.  Therefore, we fixed O/N and Ar/N to those expected by NRLMSIS~2.0 \cite{2021ESS....801321E}.  Abundances of other elements are fixed to be zero.  This treatment is the same as in our previous paper \cite{2021JGRA..12628886K,2023JGRA..12830797K}.  

We fitted the X-ray spectra extracted from altitude layers of either 5\,km or 2\,km spacing below 150\,km altitude, depending on the photon statistics.  Figure~\ref{fig:spec} right exhibits example best-fit models with the data for occultation ID~10.  The four datasets from different altitude layers of 110--115\,km, 100-105\,km, 95--100\,km, and 90--95\,km are well fitted with this model.  In this way, we obtained a total N+O (molecules and atoms are combined) column density at each altitude layer.  Finally, we calculated the local number density by inverting the Abel integral equation, following the literature \cite{1972P&SS...20.1727R,2021JGRA..12628886K,2023JGRA..12830797K}.  

Figure~\ref{fig:density_r} (a)--(d) and Fig.~\ref{fig:lospos_densityR} lower panel show resultant N $+$ N$_2$ $+$ O densities ($n_{\rm{N}} + n_{\rm{N_2}} + n_{\rm{O}}$) as a function of tangent altitude for all the occultations.  Here, the densities measured are divided by those expected by either GAIA or NRLMSIS~2.0 models to consider spatiotemporal BG density variations.  GAIA (Ground-to-topside model of Atmosphere and Ionosphere for Aeronomy) is a whole atmosphere-ionosphere coupled model.  Detailed description of GAIA is provided by \citeA{2011JGRA..116.1316J,2012JGRA..11710323J}.  In the present study, we used data obtained from the realtime version of GAIA \cite{2020EP&S...72..178T}.  GAIA incorporates the meteorological reanalysis data JRA-55 \cite{2015JMeSJ..93....5K} below 40\,km altitude through a nudging method, which enables us to simulate day-to-day variations in the atmosphere and the ionosphere. The effect of the Tonga eruption is not included in the GAIA simulation. This version of GAIA assumes a quiet convection electric field with a cross polar cap potential of 30\,kV.  Therefore, the effect of geomagnetic disturbances is also not included in the GAIA data.  For comparison, we also compared our data with the NRLMSIS~2.0 model \cite{2021ESS....801321E}.  The NRLMSIS~2.0 model is calculated at appropriate conditions for each occultation given in Table~\ref{tab:obs_list}, together with Ap and F10.7 indices taken from the web site of the GFZ Potsotom ({\tt https://www.gfz-potsdam.de/en/about-us/organisation}).  As with the GAIA model, the NRLMSIS~2.0 model does not take into account the effect of the eruption.  Gray areas in Fig.~\ref{fig:density_r} illustrate the standard deviation around the mean of the four data groups on 14 January (A--D in Table~\ref{tab:obs_list}).  Other data with filled circles are responsible for individual occultations on 15 January: IDs~10, 11, and 12 in the left panel, and 13, 14, and 15 in the right panel.  As expected from the light curves in Fig.~\ref{fig:LC}, the densities are generally more depleted after the HTHH eruption than before.  The density depletion is particularly severe for occultation ID~10, which is closest to HTHH in both time and location.  In addition, the wave-like density perturbations are clearly seen in the post-eruption data.


\section{Discussion}

We have found a large-scale neutral density depletion as well as density waves in the MLT (90--150\,km altitude) around the HTHH volcanic eruption center, based on atmospheric X-ray occultations of the Cas~A supernova remnant observed with Insight-HXMT/LE.  Here, we compare our results with other earlier HTHH observations to understand the nature of the neutral density disturbances observed in the MLT.

\begin{figure}
\includegraphics[scale=1.0]{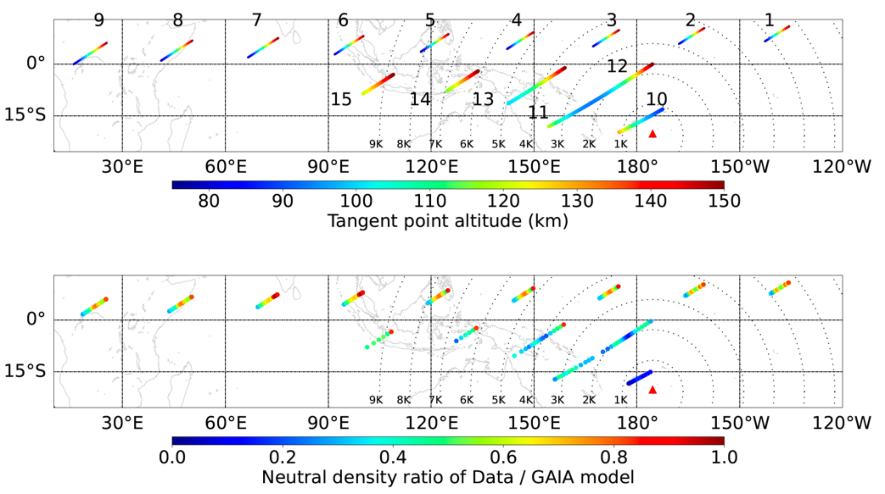}\\
\caption{Upper panel: Tangent point altitudes during the atmospheric occultations analyzed in this paper, overlaid on the Earth's surface map.  The thin and thick lines are responsible for January 14 and 15 in 2022, respectively.  The HTHH location is marked as a red triangle.  The iso-distance circles from the eruption epicenter are shown in black dashed lines (unit: km).  Lower panel: Density ratios of our observations to the GAIA model.  Occultation IDs~1+2, 3+4+5, and 6+7, and 8+9 are combined together to improve the photon statistics.}
\label{fig:lospos_densityR}
\end{figure}

\begin{figure}[t]
\begin{tabular}{cc}
\centering
\includegraphics[scale=1.0]{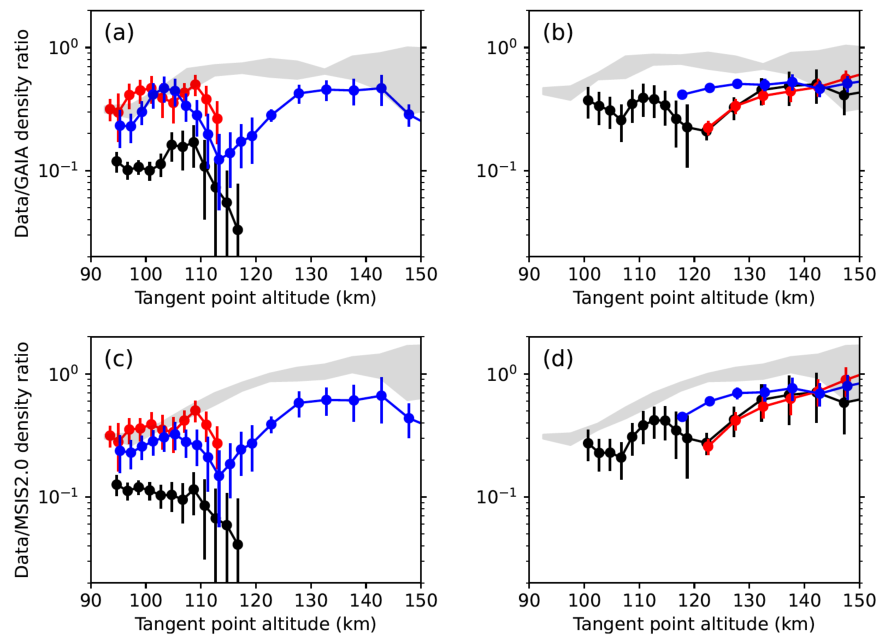}
\end{tabular}
\caption{(a) Density ratios of our measurements to the GAIA model.  The data in black, red, and blue are responsible for ID~10, 11, and 12 in Table~\ref{tab:obs_list}, respectively.  The gray area shows a standard deviation of ID~1--9.  (b) Same as left, but for ID~13, 14, and 15.  (c, d) Same as (a) and (b) but the density model is replaced to NRLMSIS~2.0.  For clarity, we shift the x-axis by $-0.3$ ($+0.3$)\,km for occultation IDs of 10 and 13 (12 and 15).}
\label{fig:density_r}
\end{figure}

From Figs.~\ref{fig:lospos_densityR} lower and \ref{fig:density_r}, significant density depletions are clear in all the post-eruption data except for the occultation ID~15.  Therefore, it is natural to conclude a density depletion within a distance of $\sim$7,000\,km from the epicenter lasting for at least 8.2\,hr after the eruption, although we cannot fully rule out the possibility that the X-ray occultations happned to capture only depleted density regions during strong density disturbances, given that our measurements are sparse in time and space.  In addition, our data seem to show a localized strong neutral depletion (by a factor of $\sim$10) extending within $\sim$1,000\,km from the epicenter, surrounded by a relatively mild depletion (by a factor of $\sim$2) extending up to $\sim$7,000\,km.  Both of the localized-strong and extended-mild depletions of the neutral density can not be reproduced by the hydrostatic atmosphere-ionosphere coupled model \cite{2023EPS...75...68M}.  We also found that the non-hydrostatic atmosphere model of \citeA{2024EPS...76...15S} could not reproduce the neutral density depletion, although small density depletion appears in the lower thermosphere in the vicinity of the HTHH volcano.  It is beyond the scope of this study to solve the mechanism of the neutral atmosphere variations.

The spatial density structure, i.e., the strong depletion surrounded by a mild depletion, resembles that of the ionospheric hole detected near the HTHH eruption, i.e., a local strong depletion in the ion density (by 1--2 orders of magnitude) within $\sim$1,500\,km from the volcano and a largely extending TEC depletion within a distance of $\sim$4,000\,km from the epicenter \cite{2023JGRA..12830984H,2023EP&S...75..184C}.  \citeA{2023JGRA..12830984H} interpreted that the HTHH eruption affects only the core area ($<$1,500\,km), and that the TEC depletion seen in the surrounding area is caused by the geomagnetic storm just before the HTHH eruption.  However, it is unclear if the same discussion holds for the neutral density disturbances found in this work.  Another possibility to explain the two types of depletions would be that the strong depletion near the volcano is caused by direct eruptive explosion and the mild depletion and wavy density structures in the far field are caused by Lamb waves.

Using accelerometers aboard GRACE-FO and Swarm-C, \citeA{2023GeoRL..5002265L} measured neutral densities at $\sim$500\,km altitude during the HTHH eruption.  They found a large crater-like (a radius $\sim$ 10,000\,km), long-lasting (duration $>$ 12\,hr) neutral density depletion near the epicenter \cite{2023GeoRL..5002265L}.  The size and duration of the neutral density hole seen at $\sim$500\,km altitude are roughly consistent with those seen at $\sim$100\,km altitude.  This similarity indicates that the entire thermosphere, at least in the altitude range of 100--500\,km, underwent volcano-induced neutral density depletions over an enormous spatial extent lasting for more than $\sim$10\,hr.

\citeA{2023GeoRL..5002265L} also found three density waves at $\sim$500\,km altitude propagating concentrically across the globe with nearly unchanged speeds of 452, 304, and 207\,m\,s$^{-1}$.  Of these, the slowest wave might have been also identified in the MLT over Chilean coast 12\,hr after the eruption as observed by CONDOR \cite{2023AnGeo..41..197S}.  Therefore, we checked if there are any indications of the neutral density waves seen at $\sim$500\,km in our vertical density profiles in the MLT.  Because our occultation positions are well within the first wave, it is unlikely to detect signatures associated with these waves.  However, signatures of the second and third waves with phase speeds of 304 and 207\,m\,s$^{-1}$ might be visible in our data.  The radii of the second/third waves are calculated to be $\sim$3,700/2,500\,km, $\sim$5,500/3,700\,km, $\sim$5,600/3,800\,km, $\sim$7,200/4,900\,km, $\sim$9,000/6,100\,km, and $\sim$10,700/7,300\,km, for the six occultations at 3.4\,hr (ID~10), 5.0\,hr (ID~11), 5.1\,hr (ID~12), 6.6\,hr (ID~13), 8.2\,hr (ID~14), and 9.8\,hr (ID~15)\,hr after the HTHH eruption, respectively.  Figure~\ref{fig:lospos_densityR} shows that the earlier four X-ray occultations (IDs~10--13) are within both of the two waves.  On the other hand, occultation IDs~14 and 15 seem to capture third and second waves at $\sim$150\,km altitude, respectively, remaining some possibilities for us to detect their signatures in our data.  However, no clear signatures are seen in both the X-ray light curves in Fig.~\ref{fig:LC} right and the density ratio profiles in Fig.~\ref{fig:density_r} right.  This is not surprising.  For example, X-ray data are not very sensitive to density variations at altitude 150\,km where the atmospheric absorption is relatively weak.  Moreover, the atmospheric waves may not reach the $\sim$150\,km and $\sim$500\,km altitudes at the same time.  These effects make it difficult for us to identify the wave signatures.  

The post-eruption density profiles in Fig.~\ref{fig:density_r} exhibit wavy structures whose amplitudes are significantly larger than fluctuations seen in pre-eruption data, shown as gray areas in Fig.~\ref{fig:density_r}.  Spatial wavelengths of the density waves could be in either vertical or horizontal direction, because the tangent point moves both in vertical and horizontal directions at the same time.  If we assume that the density wave is horizontally uniform ($\lambda_{\rm{h}}\rightarrow \infty$) and vertically propagating like a gravity wave, its vertical wavelengths can be estimated from Fig.~\ref{fig:density_r} to be $\lambda_{\rm{z}} \sim\ 20$\,km.  On the other hand, if we assume that the density wave is vertically uniform ($\lambda_{\rm{z}}\rightarrow \infty$) and horizontally propagating like the Lamb wave, the horizontal wavelength is estimated from Fig.~\ref{fig:lospos_densityR} lower to be $\lambda_{\rm{h}}\sim 1,000$\,km.  Either $\lambda_{\rm{z}}\sim 20$\,km or $\lambda_{\rm{h}}\sim 1,000$\,km is consistent with those estimated for the strong and slow gravity waves observed with TIMED/SABER \cite{2022GeoRL..4998339L}.

\section{Conclusions}

We presented neutral density disturbances in the MLT just before and after the HTHH volcanic eruption on 15 January 2022, for the first time.  X-ray observations of the bright supernova remnant Cas~A with the Chinese X-ray astronomy satellite, Insight-HXMT, serendipitously captured atmospheric X-ray occultations around the HTHH volcano.  Using the data on 14 and 15 January, we derived neutral density profiles covering an altitude range of 90--150\,km and a horizontal length of $\sim$2,000\,km.  Shortly after the eruption, we found severe density depletions in the vicinity of the epicenter, by factors of $\sim$2--10 less than the control experiments estimated by the GAIA and MSIS models.  The depletion lasts for over 10\,hr, and extends up to $\sim$10,000\,km toward the west of the epicenter.  Both the duration and horizontal spatial extension of the neutral depletion are consistent with those observed at $\sim$500\,km altitude, whereas the depletion level is much larger at $\sim$100\,km than at $\sim$500\,km.  It seems that present atmosphere-ionosphere coupled models of the HTHH volcanic eruption are unable to reproduce the observed neutral density variations driven by the eruption, indicating that further modeling investigation is necessary.  In addition, the neutral density profiles suggest a typical length scale of either $\sim$20\,km (vertical) or $\sim$1,000\,km (horizontal). This may be caused by Lamb waves or gravity waves triggered by the volcanic eruption.  We finally remark that the X-ray occultation technique will become more and more important to explore the MLT which is an important but least studied atmospheric layer.

\section{Open Research}
All the data used in this paper, software, and the calibration database of the Insight-HXMT satellite are available at: \url{http://archive.hxmt.cn/proposal}.  The XSPEC and some other softwares used in our spectral analysis are available at: \url{https://heasarc.gsfc.nasa.gov/docs/software/lheasoft/}.  


\acknowledgments

We thank Mingyu Ge and Xiaobo Li for their kind assistance in the analysis of the Insight-HXMT data.  We are also grateful to Drs. Tsubasa Tamba and Kazuo Shiokawa for advices on the analysis of Centaurus X-3, a target observed with NuSTAR from 12 January to 16 January in 2022, and insights into the gravity waves, respectively.  This work was supported by the Japan Society for the Promotion of Science KAKENHI grant numbers 20K20935 (SK and MST), 21H01150 (Yasunobu Miyoshi, HJ, CT, HS, and HF), and 19K03942 (CT, Yasunobu Miyoshi, and HF).


%




 \begin{table}
 \caption{Information about X-Ray Occultations at Tangent Altitudes of $\sim$100\,km}
 \centering
 \begin{tabular}{lccccccccc}
 \hline
  Universal Time$^a$ & Local Time & Lat. & Lon. & Alt. & $\Delta t$$^b$ & Dist.$^c$ & Type & ID & Group \\
   & (hr) & ($^\circ$) & ($^\circ$) & (km) & (hr)& (km)&  & &  \\
 \hline
    2022-01-14 08:01:04 & 22.672 & 8.177 & 219.813 & 100   & $-20.2$& 5,080  & Rising & 1  & A  \\
    2022-01-14 09:36:06 & 22.591 & 7.518 & 194.844 & 100   & $-18.6$& 3,364  & Rising & 2  & A  \\
    2022-01-14 11:11:08 & 22.511 & 6.852 & 169.884 & 100   & $-17.1$& 3,496  & Rising & 3  & B  \\
    2022-01-14 12:46:10 & 22.431 & 6.191 & 144.925 & 100   & $-15.5$& 5,338  & Rising & 4  & B  \\
    2022-01-14 14:21:12 & 22.351 & 5.521 & 119.970 & 100   & $-13.9$& 7,746  & Rising & 5  & B  \\
    2022-01-14 15:56:11 & 22.260 & 4.749 & 94.854 & 100    & $-12.3$& 10,331 & Rising & 6  & C  \\
    2022-01-14 17:31:11 & 22.174 & 4.014 & 69.809 & 100    & $-10.7$& 12,951 & Rising & 7  & C  \\
    2022-01-14 19:06:09 & 22.079 & 3.204 & 44.647 & 100    & $-9.1$ & 15,521 & Rising & 8  & D \\
    2022-01-14 20:41:07 & 21.984 & 2.390 & 19.488 & 100    & $-7.6$ & 17,718  & Rising & 9  & D \\
    2022-01-15 07:39:29 & 19.804 & $-15.991$ & 182.188 & 100 & $+3.4$ & 571    & Setting& 10 & E \\
    2022-01-15 09:15:29 & 19.960 & $-14.785$ & 160.533 & 100 & $+5.0$ & 1,945  & Setting& 11 & F \\
    2022-01-15 09:19:13 & 20.854 & $-7.441$ & 173.013 & 100  & $+5.1$ & 2,668  & Rising & 12 & G \\
    2022-01-15 10:52:33 & 20.372 & $-11.487$ & 142.448 & 100 & $+6.6$ & 4,677  & Rising & 13 & H \\
    2022-01-15 12:28:49 & 20.586 & $-9.670$ & 121.580 & 110  & $+8.2$ & 6,945  & Rising & 14 & I \\
    2022-01-15 14:02:30 & 20.234 & $-12.926$ & 92.885 & 115  & $+9.8$ & 9,840  & Rising & 15 & J \\
 \hline
 \multicolumn{10}{l}{{\tiny{$^{a}$In format of YYYY-MM-DD HH:MM:SS.  $^b$Time after the HTHH eruption.  $^c$Horizontal distance from the HTHH volcano.}}}
 \end{tabular}
\label{tab:obs_list}
\end{table}

\end{document}